\documentclass[amsmath,amssymb,aps,prl,reprint]{revtex4-2}
\usepackage{graphicx}
\usepackage{dcolumn}
\usepackage{bm}
\usepackage[normalem]{ulem}
\usepackage{upgreek}
\usepackage{xcolor}
\usepackage{mathrsfs}
\usepackage[sort&compress]{natbib}

\begin{document}
\title{D’yakov-Kontorovich instability of expanding shock waves}
\author{César Huete}
\email{chuete@ing.uc3m.es}
\affiliation{Grupo de Mec\'anica de Fluidos. Universidad Carlos III de Madrid, Legan\'es, Spain}

\author{Alexander L. Velikovich}
\affiliation{Plasma Physics Division, Naval Research Laboratory, Washington, DC 20375, USA}

\date{\today} 
\begin{abstract}
In the range of $h_c<h<1+2 M_2$, where $h$ is the D’yakov-Kontorovich parameter, $h_c$ is its critical value corresponding to the onset of the spontaneous acoustic emission, and $M_2$ is the downstream Mach number, the classic analysis predicts a special form of the instability of isolated steady planar shock waves: non-decaying oscillations of shock-front ripples. For spherically and cylindrically expanding steady shock waves, we demonstrate instead an instability in a literal sense, a power-law growth of shock-front perturbations with time. As the parameter $h$ increases from $h_c$ to $1+2M_2$, the instability power index grows from zero to infinity. Shock divergence is a stabilizing factor, and instability is found for high angular mode numbers.
\end{abstract}

\maketitle

The shock stability problem is one of the fundamental problems of compressible fluid dynamics. Despite the extensive literature accumulated since the pioneering works of \citet{DYakov1954} and \citet{Kontorovich1957} (DK) in the 1950s, the problem is not fully resolved, particularly when realistic boundary conditions are considered.

For a shock wave to be evolutionary, which is a pre-requisite for its stability \citep{Landau1987}, the upstream and downstream Mach numbers in the shock-stationary reference frame must satisfy $M_1>1$ (supersonic upstream) and $M_2<1$ (subsonic downstream), respectively. The shock front is acoustically coupled with downstream influences. The inclusion of a supporting mechanism, which is, in fact, a necessary condition for the shock to be steady, affects the shock behavior, and ultimately, its stability limits. If the shock front is under-supported (followed by an expansion wave, gradually reducing its strength) or over-supported (followed by a compression wave, gradually increasing its strength), the stability analysis applies to the whole flow. It can be unstable in either case, even when the shock front \textit{per se} is surely stable (blast wave \citep{Ryu1987,Grun1991,Sanz2016} and converging shock \citep{Gardner1982,Murakami2015} in an ideal gas are examples). When we focus on studying the shock front’s stability, it must be steady, which implies a piston maintaining a constant pressure behind it.

\begin{figure}
 \centering
  \includegraphics[width=0.47\textwidth]{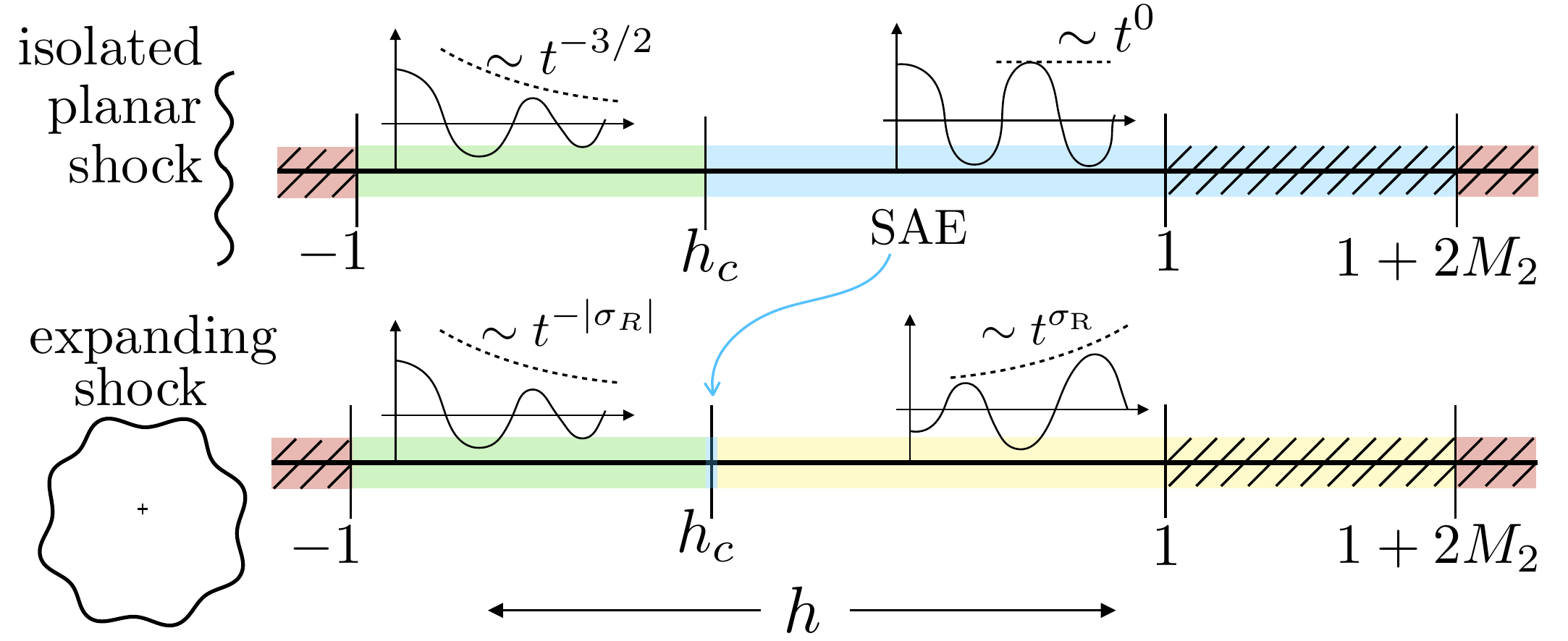}
  \caption{Distinguished regimes for isolated planar shocks and expanding accretion shocks along the variable $h$.}
  \label{fig1}
\end{figure}

The stability conditions for a steady isolated shock wave can be written in terms of the DK parameter
\begin{equation}
h= \frac{p_2-p_1}{V_1-V_2}\left(\frac{{\rm d} V_2}{{\rm d}  p_2}\right)_{\hspace{-0.1cm}H}=-u_2^2\left(\dfrac{\partial \rho_2}{\partial p_2}\right)_{\hspace{-0.1cm}H}
\label{h}
\end{equation}
that measures the slope of the Hugoniot curve relative to the Rayleigh-Michelson line on the ($V,p$) plane. Here $p$, $\rho$, $V=1/\rho$, and $u$ denote the pressure, density, specific volume, and fluid velocity with respect to the shock front, respectively, subscripts 1 and 2 refer to pre- and post-shock states, and the derivatives are calculated along the Hugoniot curve with the pre-shock pressure and density fixed. For an isolated steady planar shock front, the classic stability theory predicts an oscillatory decay of perturbations as $t^{-3/2}$ ($t^{-1/2}$ in the strong-shock limit), with a constant oscillation frequency, for any wavenumber (see \citep{Roberts1945} for an ideal-gas equation of state (EoS) and \citep{Bates2004} for an arbitrary EoS), provided that the parameter $h$ is in the stable range, $-1<h<h_c$, where 
\begin{equation}
h_c = \dfrac{1-M_2^2\left(1+R\right)}{1-M_2^2\left(1-R\right)},
\label{hc}
\end{equation}
and $R=\rho_2/\rho_1$ is the shock density compression ratio. For an ideal-gas EoS, $h=-1/M_1^2$, $h_c = -1/(2M_1^2-1)$, and the stability conditions are always satisfied. For $h_c<h<1+2M_2$, shock perturbations with certain wavevectors oscillate at constant amplitude, causing spontaneous acoustic emission (SAE) from the shock front \citep{Kontorovich1957,Landau1987}. As noted by \citet{Landau1987}, \S 90, SAE is not an ``\textit{instability in a literal sense}" as there is no real growth. Absolutely unstable ranges are $h<-1$ and $h>1+2M_2$, for which the exponential growth of shock-front perturbations is associated with a shock breakup into several simple waves \citep{Kuznetsov1989,Menikoff1989}. These stability limits are sketched in Fig.~\ref{fig1} (top), where the hatched regions correspond to conditions that render multi-valued \citep{Erpenbeck1962,Kuznetsov1984} or multi-wave \citep{Kuznetsov1989,Menikoff1989} solutions of the planar Riemann/piston problem.

The presence of a rigid piston supporting a planar shock enables acoustic waves to reverberate between the shock front and the piston. This effect does not qualitatively change the shock-front perturbation behavior in the absolutely stable, $-1<h<h_c$ \citep{Freeman1955,Zaidel1960,Wouchuk2004,Bates2012,Bates2015,Fowles1973}, and unstable, $h<-1$ and $h>1+2 M_2$, parameter ranges. But it can make a difference in the marginally stable/SAE range, $h_c<h<1+2M_2$, although there is no consensus in the literature about the piston’s effect. As noted in Refs.\citep{Fowles1973,Kuznetsov1984}, normally incident acoustic waves are amplified upon reflection from the shock front at $h>1$. Then the amplitude of a reverberating acoustic wave grows as a power of time for $h>1$ (see below), so the whole hatched area in the top part of Fig.~\ref{fig1} becomes unstable. The stability analysis of the initial-value problem in Ref.\citep{Wouchuk2004} for $h_c<h<1-2M_2^2$ did not find any qualitative distinctness in the shock front perturbation behavior when a piston is involved. On the other hand, Ref.\citep{Bates2015}, found instability, a linear growth of shock perturbations in the whole range $h_c<h<1+2M_2$.

A spherically or cylindrically expanding shock front, just like a planar one, can be steady if a constant-velocity inflow of incident mass supports it. Then the shocked mass is uniform and at rest, and the center or axis of symmetry plays the role of a rigid resting piston. Such an expanding accretion-shock flow is described by a self-similar solution of ideal compressible fluid dynamics equations, which exists for an arbitrary EoS of the fluid and the parameters of the incident flow \citep{Velikovich2018}. It is labeled \textit{generalized Noh solution}, as opposite to the classic solution \citep{Noh1987} constructed for an ideal gas EoS in the strong-shock limit. Stability analysis of this solution can be carried out in the same way as done for the classic Noh solution in Ref.\citep{Velikovich2016}. Details of this derivation will be published separately. Here we present the results on the DK instability of expanding shock waves. 

The expanding shock flow is governed by the continuity equation, the Euler equation, and the adiabaticity equation. They are supplemented with the EoS specifying the dependence of specific energy on pressure and density, $E=E(p,\rho)$. The EoS determines the dependence of the sound speed, $c=\sqrt{(\partial p/\partial \rho)_S}$, where $S$ is the entropy, on the pressure and density, as well as the Hugoniot conditions. Our spherically or cylindrically-symmetric unperturbed flow is initialized as follows. 
At $t=0^-$, the fluid has uniform density $\rho_0$, pressure $p_0$, and velocity $v_0$ directed to the center or axis of symmetry. At $t=0^+$, an accretion shock emerges from the origin. It expands at constant speed $u_2=M_2 c_2$ and puts the downstream flow at rest, with uniform post-shock density $\rho_2$, pressure $p_2$, and sound speed $c_2$. The converging fluid is adiabatically compressed, from its initial density $\rho_0$, pressure $p_0$, and sound speed $c_0$ at infinity to $\rho_1$, $p_1$, and $c_1$, respectively, ahead of the shock front. The flow is self-similar, all flow variables depend on the time $t$ and radius $r$ via the normalized coordinate $\xi=r/(u_2 t)$. Calculating the self-similar profiles as explained in \citep{Velikovich2018}, one finds the velocity $u_2$ of the expanding shock front, the pre- and post-shock parameters.

Figure 2 (top) demonstrates self-similar profiles calculated for a spherically expanding shock wave in a non-ideal gas with the van der Waals (vdW) EoS defined by
\begin{equation}
E(p,\rho)=\dfrac{(p+a\rho^2)(1-b \rho)}{\rho(\gamma-1)}  -a\rho.
\label{E_vdw}
\end{equation}

\begin{figure}
 \centering
    \includegraphics[width=0.4\textwidth]{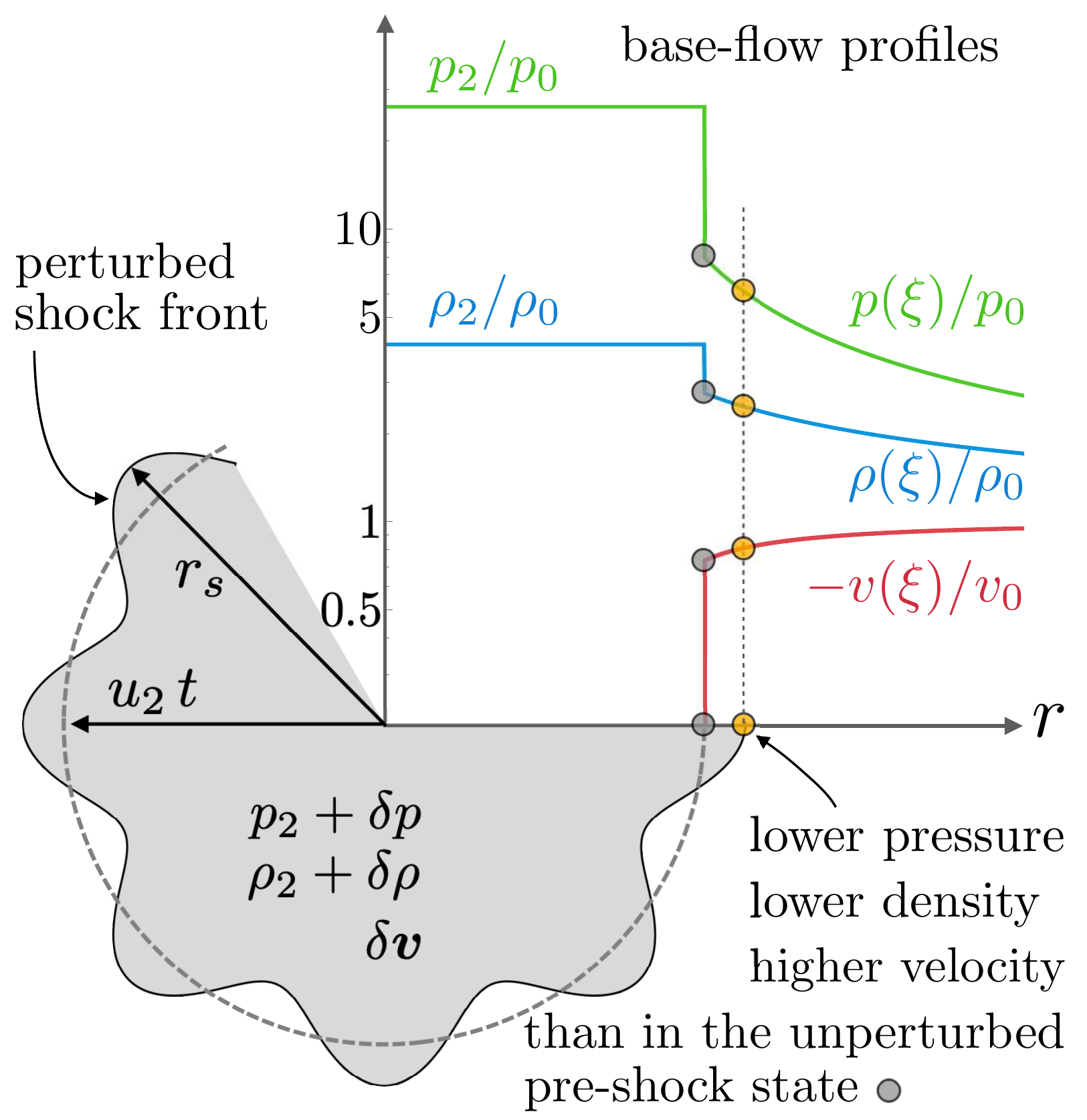}
  \caption{Sketch of a spherical shock steadily expanding through the non-uniform converging isentropic flow and perturbed with $\nu=3$ and $j=8$.}
  \label{fig2}
\end{figure}

In addition to the dimensionless adiabatic exponent, $\gamma>1$, this EoS contains two positive dimensional parameters, $a$ and $b$, which account for intermolecular attraction and the volume occupied by the molecules, respectively. Note that the EoS \eqref{E_vdw} cannot be presented in the reduced Mie-Gr{\"u}neisen form $E(p,\rho)=p\,f(\rho)$, where $f(\rho)$ is an arbitrary positive function of density, making it possible to construct classic self-similar solutions, including converging-shock and blast-wave solutions \citep{Sedov1993}. However, as explained in \citep{Velikovich2018}, generalized Noh solutions can be obtained for arbitrary EoS. The profiles shown at the top of Fig.~\ref{fig2} have been constructed for the vdW EoS parameters $\gamma=5/3$, $a\rho_0^2/p_0=1$, $b\rho_0=1/10$, $\nu=3$, and $v_0/c_0=3/2$.

Now consider a small-amplitude perturbation of such flow. For a spherical symmetry, a three-dimensional shock-front shape perturbation is expanded into spherical functions:
\begin{equation}
\dfrac{r_s (\theta,\phi,t)}{u_2 t}-1= \epsilon\sum_{l=0}^\infty \sum_{m=-l}^l \zeta_{l,m}\left(\frac{t}{t_0}\right)^{\hspace{-0.1cm}\sigma_{l,m}} \hspace{-0.4cm} P_l^m(\cos\theta)e^{{\rm i}m\phi},
\label{rs}
\end{equation}
where $\epsilon$ is a small parameter, $t_0$ is the time scale, $l$ and $m$ are polar and azimuthal mode numbers, $P_l^m(z)$ is associated Legendre function, $\zeta_{l,m}$ is the complex amplitude of the $(l,m)$ mode, and $\sigma_{l,m}$ is a complex dimensionless eigenvalue. The shape \eqref{rs} is scale-free, which makes it possible to separate the time variable in the perturbed fluid equations, seeking their solution in the form
\begin{equation}
\begin{bmatrix}
\delta \rho/\rho_2 \\
\delta p/p_2\\
\delta v_r/u_2\\
\delta d/u_2
\hspace{-0.15cm}
\end{bmatrix}
  \hspace{-0.1cm}= \epsilon\sum_{l=0}^\infty \sum_{m=-l}^l\begin{bmatrix}
G_{l,m} (\xi) \\
P_{l,m} (\xi)\\
V_{l,m} (\xi)\\
D_{l,m} (\xi)
\end{bmatrix}
\hspace{-0.15cm}
\left(\frac{t}{t_0}\right)^{\hspace{-0.1cm}\sigma_{l,m}} \hspace{-0.4cm} P_l^m(\cos\theta)e^{{\rm i}m\phi},
\label{perturbation}
\end{equation}
where the transverse velocity divergence function is $\delta d=r\nabla_\perp\cdot \delta \boldsymbol{v}$. The dimensionless eigenfunctions $G_{l,m}(\xi)$, $P_{l,m}(\xi)$, $V_{l,m}(\xi)$, and $D_{l,m}(\xi)$ are sought in the range from $\xi=0$ (the origin) to $\xi=1$ (the shock front). The perturbed Hugoniot conditions \citep{Landau1987} are translated into the boundary conditions for the eigenfunctions at $\xi=1$, which are supplemented by the requirement of the regularity of the solution at $\xi=0$. As in \citep{DYakov1954,Kontorovich1957,Landau1987}, the pre-shock flow is assumed to be unperturbed. 

Our analysis is conventionally done one Fourier-Legendre mode at a time, so we omit the mode-number subscript, denoting $\sigma_{l,m}=\sigma=\sigma_{\rm R}+ i \sigma_{\rm I}$. The real part, $\sigma_{\rm R}$, determines if the relative shock ripple amplitude \eqref{rs} and the absolute perturbation amplitudes of density, pressure, and velocity \eqref{perturbation} decrease or grow as a power of time. The latter case of $\sigma_{\rm R}>0$ corresponds to an instability. The imaginary part, $\sigma_{\rm I}$, determines the time-dependent oscillation frequency of these perturbations. The stability analysis can be done similarly for cylindrically expanding shocks, but with the following restriction. Only the two-dimensional filamentation perturbations $\sim \exp(i m \phi)$ are scale-free, thereby enabling separation of variables in the perturbation equations \citep{Velikovich2016}. Then the double sum over $l$ and $m$ in \eqref{rs} and \eqref{perturbation} is replaced with a single sum over $m$ from $0$ to infinity.

The stability analysis for both cases \citep{Velikovich2016} can be summarized in a dispersion equation:
\begin{widetext}
\begin{equation}
\Big\{\left(\sigma+\nu-1\right) \Big[R (\nu-1)-\sigma-\nu \Big] + R\, j (j+\nu-2)\Big\}(1+h)F_{1s}^+
+\Big[2(\sigma+\nu)-R(\nu-1)\left(1 + h_1 \right)\Big](\sigma+\nu+j-1)F_{1s}^-=0.
\label{diserpsion}
\end{equation}
\end{widetext}

Here, the geometric parameter $\nu=2$ and $3$ for cylindrical and spherical symmetry, respectively. The angular mode number $j$ stands for $m$ in the former case and $l$ in the latter, when the eigenvalues depend on the polar mode number only. The functions
\begin{equation}
F_{1s}^{\pm}= {}_2 F_1\left(\dfrac{j-\sigma}{2},\dfrac{j\pm1-\sigma}{2};j+\dfrac{\nu}{2};M_2^2\right)
\label{F1pm}
\end{equation}
are Gauss hypergeometric functions that can be reduced to associated Legendre functions. Finally, the parameter
\begin{equation}
 h_1= \dfrac{h}{M_{1}^2-1}\left[\dfrac{1}{c_{1}^2}\left(\dfrac{\partial p_2}{\partial \rho_{1}}\right)_{\hspace{-0.15cm}\rho_2,p_{1}}\hspace{-0.4cm}+\left(\dfrac{\partial p_2}{\partial p_{1}}\right)_{\hspace{-0.15cm}\rho_2,\rho_{1}}\right]
\label{hs}
\end{equation}
accounts for the non-uniformity of the pre-shock flow. Here, the derivatives of the post-shock pressure are taken along the Hugoniot surface (rather than curve, because both pre- and post-shock parameters are varied) at constant post-shock density. For the ideal-gas EoS, we have $h_1=(\gamma-1)(1-M_1^{-2})/(\gamma+1)$. The pre-shock flow is unperturbed, but a distortion of the shock-front shape changes the pre-shock state, as illustrated in Fig.~\ref{fig2}. 

The dispersion equation \eqref{diserpsion} is a spherical/cylindrical counterpart of the DK dispersion equation for an isolated planar shock, Eq.(90.10) of \citep{Landau1987}. In planar geometry, it is impossible to derive a dispersion equation that takes a piston into account. This is why the shock-front stability analysis had either to be done heuristically \citep{Fowles1973,Kuznetsov1984} or use much more complicated mathematics \citep{Wouchuk2004,Bates2015}. By contrast, our dispersion equation \eqref{diserpsion} accounts for the piston represented by the center or axis of symmetry. For a given geometric parameter $\nu$ and four shock parameters, $M_2$, $R$, $h$, and $h_1$, the expanding shock front is unstable if for any angular mode number $j$ there is an eigenvalue with $\sigma_{\rm R}>0$. Notice that all the parameters entering \eqref{diserpsion} are real, thereby providing pairs of physically equivalent complex-conjugate eigenvalues, of which we only show those with non-negative $\sigma_{\rm I}$.   

Consider first the short-wavelength limiting regime corresponding to $\sigma_{\rm I}\rightarrow\infty$,  $\sigma_{\rm R}$ and $j$ finite. We derive from \eqref{diserpsion}
\begin{subequations}
\begin{align}
&&\sigma_{\rm R}^{(n\gg j)}\sim - \dfrac{\nu-1}{2} + \dfrac{\ln |\mathscr{R}_s|}{\ln \mathscr{D}_s}, \label{sigmaRngg1}\\
&&\sigma_{\rm I}^{(n\gg j)}\sim \dfrac{2\pi n}{\ln \mathscr{D}_s} +O(1).
\label{sigmaIngg1}
\end{align}
\end{subequations}
Here, the positive integer $n$ is the radial mode number, roughly corresponding to the number of zeroes of the pressure eigenfunction $P_{l,m}(\xi)$ in the interval $0<\xi<1$; $\mathscr{D}_s =(1+M_2)/(1-M_2)>1$ is the Doppler shift factor; and
\begin{equation}
\mathscr{R}_s = \dfrac{2M_2 -1+ h}{2M_2+1-h}
\label{reflection}
\end{equation}
stands for the reflection coefficient for an acoustic wave normally incident on the shock front from downstream. In this limit, acoustic waves reverberate almost normally to the shock front. The relevant length scale, $\sim u_2 t/n$, is much smaller than those associated with the pre-shock non-uniformity and the angular mode number, $\sim u_2 t$ and $\sim u_2 t/j$, respectively, which explains why parameters $h_1$ and $j$ do not enter Eqs.~\eqref{sigmaRngg1} and \eqref{sigmaIngg1}. Although \eqref{diserpsion} does not apply to planar geometry, $\nu=1$, the asymptotic formulas \eqref{sigmaRngg1} and \eqref{sigmaIngg1} are valid in this case, too. They describe an acoustic wave reverberating between the shock front and the piston at the speed of sound, $c_2$, whereas the shock front moves away from the piston at the velocity $u_2$. Its back-and-forth cycles increase in duration as powers of the Doppler shift factor: $t_1$, $\mathscr{D}_s t_1$, $\mathscr{D}_s^2 t_1$, ..., cf. Fig.3 of \citep{Fowles1973}. Assuming the reflection coefficient from the piston to be unity, in planar geometry each cycle multiplies the acoustic wave’s amplitude by the shock reflection coefficient: $1$, $\mathscr{R}_s$, $\mathscr{R}_s^2$, ... The amplitude thus varies as a complex power of time, the real part of the power index for $\nu=1$ being given by the second term in the right-hand side of \eqref{sigmaRngg1}. The first term, negative for $\nu=2$ and $3$, describes the attenuation of diverging acoustic waves, as explained in \citep{Velikovich2016}. The stabilizing effect of divergence is obviously stronger for spherical expansion.

For $1<h<1+2M_2$, we have  $\mathscr{R}_s>1$, so acoustic waves are amplified upon reflection from the shock front, indicating instability for planar geometry, in agreement with \citep{Fowles1973,Kuznetsov1984}. The unstable range for all cases, $\nu=1$, $2$, and $3$, is
\begin{equation}
1+2M_2\dfrac{\mathscr{D}_s^{\frac{\nu-1}{2}}-1}{\mathscr{D}_s^{\frac{\nu-1}{2}}+1}<h<1+2M_2.
\label{limits}
\end{equation}

As $h$ increases from the lower to the upper boundary of the range \eqref{limits}, the corresponding power index $\sigma_{\rm R}$ given by \eqref{sigmaRngg1} increases from zero to infinity. 

\begin{figure}
 \centering
  \includegraphics[width=0.475\textwidth]{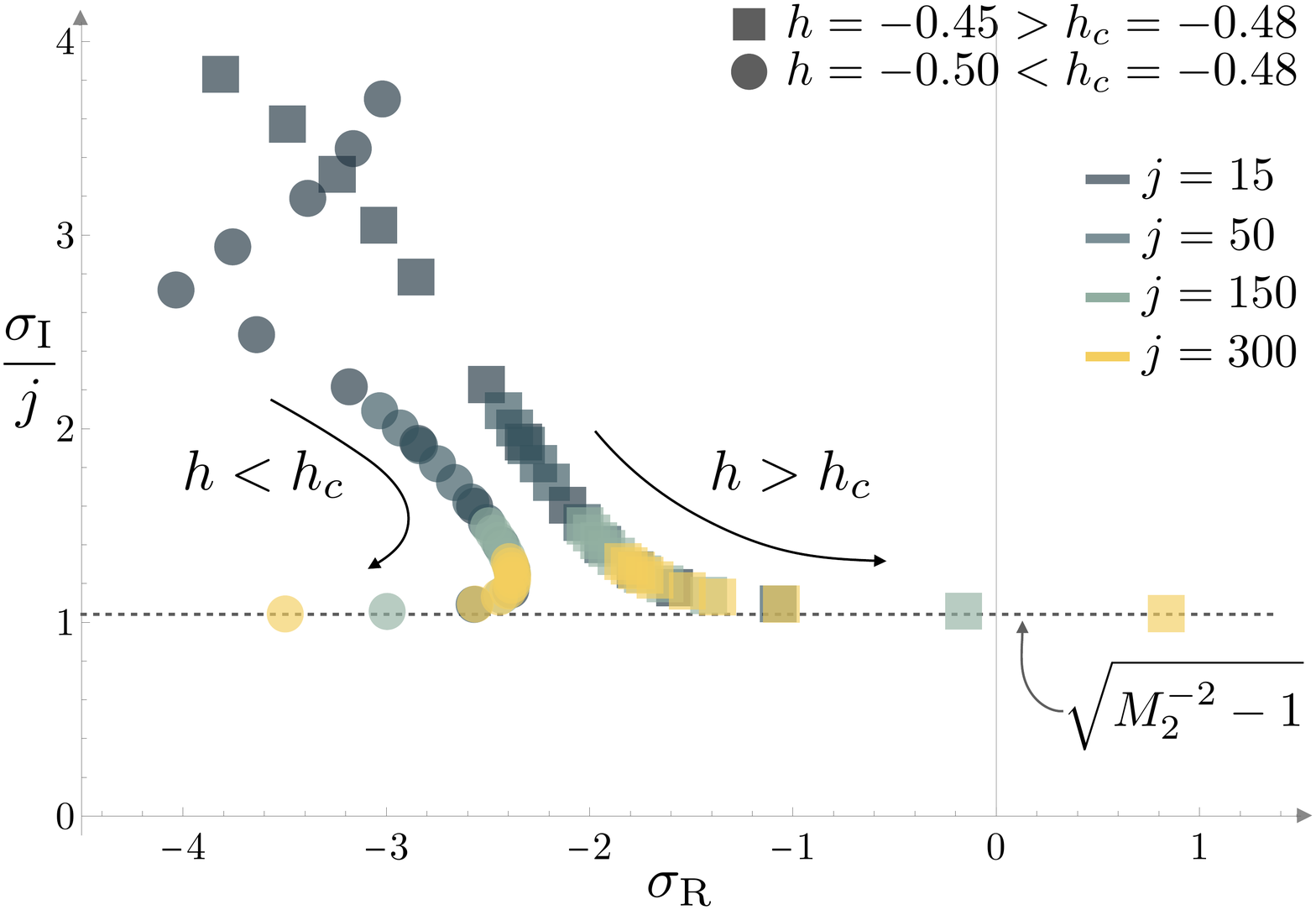}
  \caption{Eigenvalues for $\nu=2$, $R=3$, $M_2=0.7$, $h_1=0.1$, two different values of $h=-0.5$ (circles) and $h=-0.45$ (boxes), and different mode numbers $j=15$, $50$, $150$ and $300$.}
  \label{fig3}
\end{figure}

Note, however, that \eqref{limits} gives sufficient rather than necessary instability conditions. The modes with large $n\gg j$ are not necessarily the most unstable. The instability is quite possible when the right-hand side of \eqref{sigmaRngg1} is negative. To demonstrate this, we choose shock parameters arbitrarily as $\nu=2$, $R=3$, $M_2=0.7$, and $h_1=0.1$. From \eqref{hc}, we find the corresponding critical value of $h_c=-0.48$, and pick two negative values of $h=-0.45$  and $-0.5$ outside of the range \eqref{limits}. They are close, one above and the other below $h_c$. Assuming cylindrical symmetry, we calculate the eigenvalues for mode numbers $j=15$, $50$, $150$ and $300$. To show all spectra on the same scale, the vertical axis in Fig.~\ref{fig3} is normalized with respect to the angular mode number, plotting  $\sigma_{\rm I}/j$. For each $j$, Fig.~\ref{fig3} shows several complex eigenvalues corresponding to low radial mode numbers $n$, or, which is the same, low nonzero $\sigma_{\rm I}$. In the short-wavelength limit $j\rightarrow\infty$, the shock front is nearly planar, and the oscillation frequency $\omega$ of shock-front ripples is related to the wavenumber $k$ by $\omega=k c_2 \sqrt{1-M_2^2}$ \citep{Zaidel1960}. Substituting $\omega=\sigma_{\rm I}/t$ and $k=j/(u_2 t)$ for the frequency and wavenumber, respectively, we find $\sigma_{\rm I}/j=\sqrt{M_2^{}\hspace{-0.5mm}{}^{-2}-1}$, the asymptotic value shown by the horizontal dotted line. The eigenvalues computed for $h=-0.5$ (circles) are in the negative half-plane $\sigma_{\rm R}<0$, indicating stability. For a slightly larger value of $h=-0.45$ (boxes), we find $\sigma_{\rm R}>0$ at high angular mode numbers $j\geq 170$, showing instability. The most unstable eigenvalue has the lowest radial mode number. Apparently, for $j=150$, all modes are stable, whereas for $j=300$, the instability threshold is crossed somewhere between $h=-0.5$ and $-0.45$.

The instability threshold corresponds to a purely imaginary eigenvalue, $\sigma_{\rm R}=0$. To determine its location, we solve \eqref{diserpsion} for $h$ and substitute an imaginary value of $\sigma=i s$, where $s$ is real and positive, into the result, arriving at
\begin{eqnarray}
&&\hat{h}=-1 -\dfrac{F_{1s}^-}{F_{1s}^+}\Bigg|_{\sigma=i s}\hspace{-0.2cm}\times \label{hhat}\\
&&\hspace{-0.3cm}\dfrac{\Big[2(\sigma+\nu)-R(\nu-1)\left(1 + h_1 \right)\Big](\sigma+\nu+j-1)}{\left(\sigma+\nu-1\right) \Big[R (\nu-1)-\sigma-\nu \Big] + R\, j (j+\nu-2)}\Bigg|_{\sigma=i s}\nonumber\hspace{-0.6cm}.
\end{eqnarray}

For an arbitrary value of $s$, the right-hand side of \eqref{hhat} is complex, implying that $\sigma=i s$ is not a physically meaningful eigenvalue because $h$ must be real. However, for given values of $M_2$, $R$, $h_1$, $\nu$, and $j$, the equation ${\rm Im}[\hat{h}(s)]=0$ has an infinite number of solutions $s^{(1)}$, $s^{(2)}$, ..., $s^{(n)}$, ... that are actual eigenvalues and that correspond to conditions when the eigenmodes with radial numbers $n=1$, $2$, ... cross the line $\sigma_{\rm R}=0$ as $h$ increases. Substituting them into \eqref{hhat}, we find the corresponding real values of $h=h^{(1)}$, $h^{(2)}$, ..., $h^{(n)}$, ..., which accumulate near the lower boundary of \eqref{limits} as $n\rightarrow\infty$. The lowest of these values found in the range between $-1$ and the lower boundary of the interval \eqref{limits} is the instability threshold denoted by $h_{st}(M_2,R,h_1,\nu,j)$. 

\begin{figure}
 \centering     
 \includegraphics[width=0.475\textwidth]{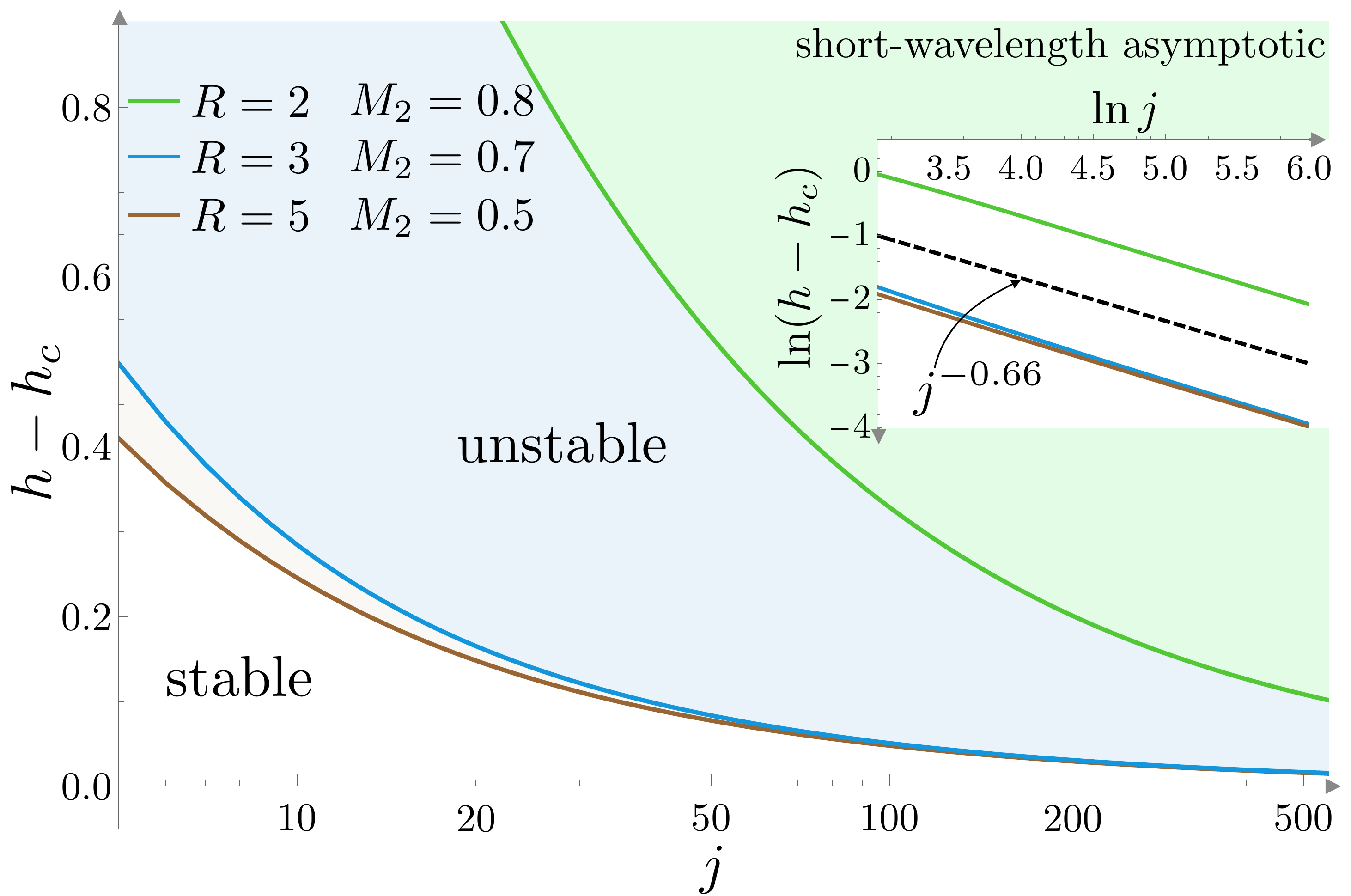}
  \caption{Instability thresholds $h_{st}-h_c$ vs. the mode number $j$ (lines), stable and unstable ranges on the plane $(j,h-h_c)$ for $\nu=2$, $h_1=0.1$ and three arbitrary sets of shock conditions.}
  \label{fig4}
\end{figure}

To compare it to the critical value for SAE given by \eqref{hc}, $h_c(M_2,R)$, we choose shock parameters arbitrarily, and plot the difference $h_{st}(M_2,R,h_1,\nu,j)-h_c(M_2,R)$ vs $j$. The results for cylindrical geometry, $M_2=0.8$, $0.7$, and $0.5$ with $R=2$, $3$, and $5$, respectively, and $h_1=0.1$ for all cases, are shown in Fig.~\ref{fig4}. Each threshold curve bounds a shaded unstable parameter range above it. For low and moderate angular mode numbers $j$, the difference is significant, manifesting the stabilizing effect of expansion. But as $j$ increases into the high-mode range, $h_{st}-h_c$ tends to zero. The inset plotted demonstrates that $h_{st}$ approaches $h_c$ at $j\rightarrow\infty$ as a negative power of the mode number, $j^{-0.66}$. To explain it, we substitute the imaginary asymptotic value $\sigma=i\,j\sqrt{M_2^{}\hspace{-0.5mm}{}^{-2}-1}$ into \eqref{hhat}. In the limit $j\rightarrow\infty$, the term in the second line tends to $-2(1-M_2^2-i M_2\sqrt{1-M_2^2})/[1+M_2^2(R-1)]$. The limiting value of the other term is
\begin{equation}
\lim_{j\rightarrow\infty}\dfrac{F_{1s}^-}{F_{1s}^+}\bigg|_{\sigma=i\,j\sqrt{M_2^{-2}-1}}=1-M_2^2+i M_2\sqrt{1-M_2^2}.
\label{limFF}
\end{equation}

Hence, in the short-wavelength limit  $j\rightarrow\infty$, for any set of shock parameters, spherical and cylindrical geometry,  $h_{st}(M_2,R,h_1,\nu,j)\rightarrow h_c(M_2,R)$, as shown in Fig.\ref{fig4}.

\begin{figure}
 \centering
    \includegraphics[width=0.475\textwidth]{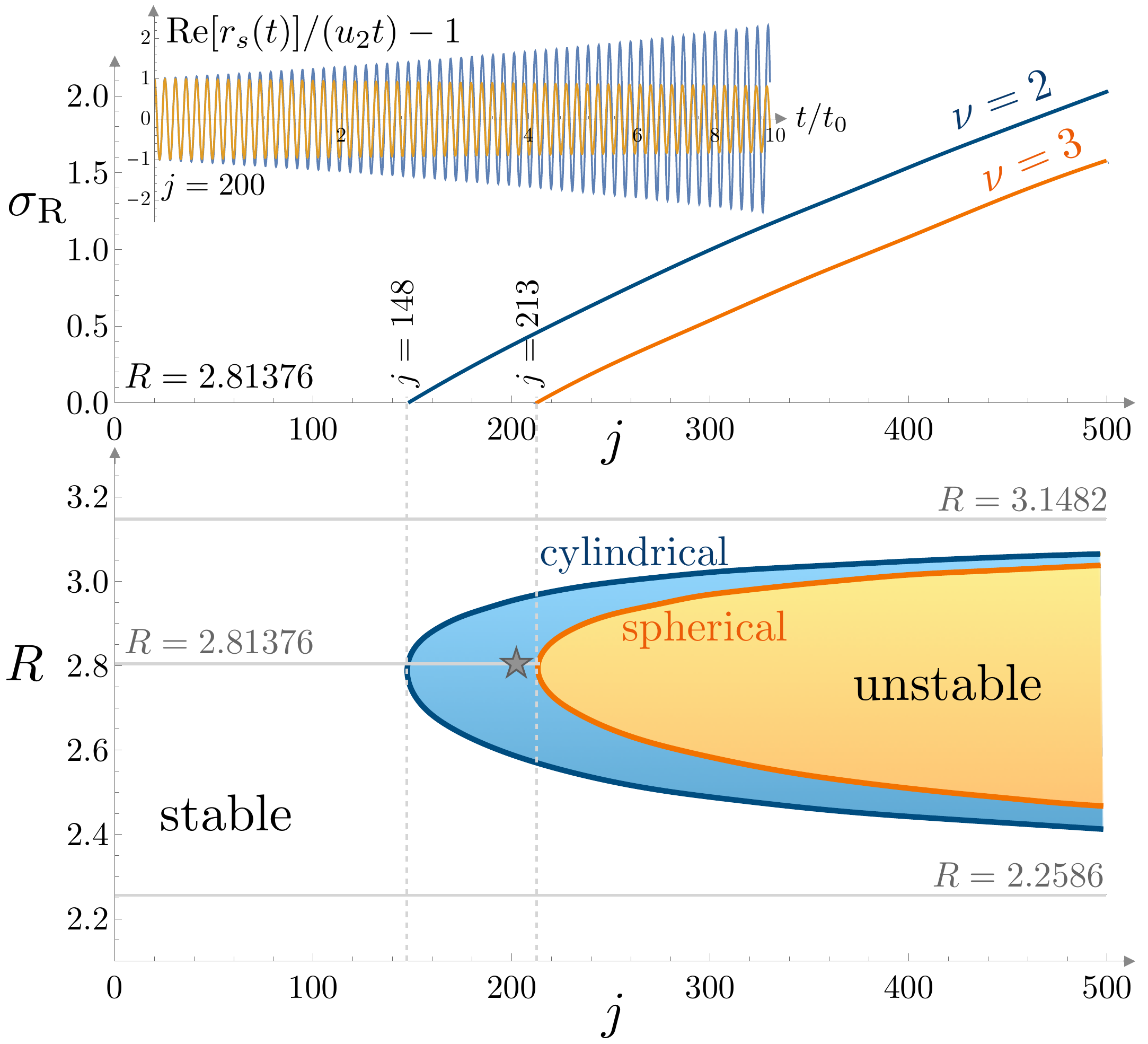}
  \caption{Stability limits and growth rate for a van der Waals EoS with $\gamma=31/30$, $a \rho_{1}^2/p_1=1/2$ and $b \rho_1=1/9$. Temporal evolution of the oscillating shock ripple amplitude for cylindrical and spherical geometries at $j=200$ (star symbol).}
  \label{fig5}
\end{figure}

Eigenmodes with high angular mode number $j\rightarrow\infty$ and low radial mode number $n$ are the most unstable. For these modes, acoustic waves reverberating behind the shock front are nearly parallel to it, in contrast with the high$-n$ modes described by \eqref{sigmaRngg1} and \eqref{sigmaIngg1}. In the short-wavelength limit $j\rightarrow\infty$, the unperturbed shock front is almost planar, and the stabilizing effect of its spherical or cylindrical expansion vanishes. Similarly, the large-scale non-uniformity of the pre-shock flow is no longer relevant. This is why the expanding shock-front instability threshold tends to the planar-geometry critical SAE value \eqref{hc}. But here, in contrast with the classic case of isolated planar shock, we have ``\textit{instability in a literal sense}" \citep{Landau1987}, all perturbation amplitudes exhibiting an oscillatory power-law growth with time, as illustrated in Fig.~\ref{fig1} (bottom) and \ref{fig5} (top).


The first example of a realistic EoS satisfying the DK instability condition was discovered by \citet{Bushman1976} near copper's liquid-vapor transition. More examples have been found since for condensed materials near the liquid-vapor transition, including water \citep{Kuznetsov1988}, a fluid approximated by the vdW EoS \citep{Bates2000}, and magnesium \citep{Lomonosov2000,Konyukhov2009}; for ionizing shock waves in inert gases \citep{Mond1994,Mond1997}; for shock waves dissociating hydrogen molecules \citep{Bates1999}; for Gbar- and Tbar-pressure range shocks in solid metals, where the shell ionization affects the shapes of Hugoniot curves \citep{Das2011,Wetta2018}; for shock fronts producing exothermic reactions, such as detonation \citep{Huete2019,Huete2020}. For all these cases, expanding shocks might be unstable. Notice that it took decades from the first identification of the unstable range for planar shocks by \citet{Bushman1976} to its numerical verification by \citet{Bates2000}. For expanding shocks, where the instability is predicted for high mode numbers, numerical simulations are very challenging since they must be capable of capturing non-dissipative high-frequency oscillations at the shock along with multidimensional acoustic perturbations in a wide range of scales. Below, we establish the instability parameter ranges and calculate the power indices for the vdW EoS, under the numerically-tested conditions employed in \citep{Bates2000}.

For the pre-shock and vdW EoS parameter values $\gamma=31/30$, $a \rho_{1}^2/p_1=1/2$ and $b \rho_1=1/9$ \citep{Bates2000}, the DK instability condition $h>h_c$ is met within an interval of shock strengths corresponding to density compressions  $2.2586<R<3.1482$. As demonstrated in Fig.~\ref{fig4}, this ensures instability of expanding shock waves for sufficiently high angular mode numbers $j$. Figure \ref{fig5} shows the stability limits for spherical and cylindrical shocks. There exists a minimum value of $j=j_{\rm min}$, below which the shock is stable for any shock compression ratio $R$. For cylindrical and spherical geometry, we have $j_{\rm min}=148$ and $213$, respectively. For the value of $R=2.81376$, which corresponds to the smallest values of $j$ in the unstable regime, Fig.~\ref{fig5} also presents the instability power indices vs $j$. As expected, the stabilizing effect of the flow divergence is stronger in spherical geometry as it demands higher angular mode numbers $j$ for the shock to become unstable. For example, the temporal evolution of the dimensionless shock perturbations \eqref{rs}, shown in Fig.~\ref{fig5} for $R=2.81376$ and $j=200$ (star symbol), predicts unstable and stable oscillations for cylindrical and spherical geometries, respectively.

To summarize, we have demonstrated that the DK instability of expanding steady shock waves drives a power-law growth of shock ripples and other flow variables' perturbations in the range $h_c<h<1+2M_2$ deemed marginally stable in the classic theory \citep{DYakov1954,Kontorovich1957,Landau1987}. The factors specific to expanding-shock flow, such as its divergence and the non-uniformity of the pre-shock profiles, do not affect the stability criteria. The difference between this case and the classic case of isolated planar shock \citep{DYakov1954,Kontorovich1957,Landau1987} is due to the piston supporting the steady shock and represented with the center or axis of symmetry.

Our conclusions are generally consistent with those of \citet{Bates2015}, who predicted a linear growth of shock perturbations for the whole range $h_c<h<1+2M_2$ in the presence of a piston. But the DK instability power indices found for spherical and cylindrical geometry vary from zero to infinity, depending on the parameter $h$ and the angular mode number $j$. 


\begin{acknowledgements}
C.H. work is produced with the support of a 2019 Leonardo Grant for Researchers and Cultural Creators, BBVA Foundation and project PID2019-108592RB-C41 (MICINN/FEDER, UE). A.L.V. work was supported by the National Nuclear Security Administration of the U.S. Department of Energy. The authors wish to thank Dr. J. W. Bates, Prof. J. G. Wouchuk, and Prof. D. Martínez-Ruiz  for carefully reading the manuscript and for their valuable comments.
\end{acknowledgements}

\end{document}